\documentclass[
aps,
prd,
nofootinbib,
superscriptaddress,
longbibliography,
floatfix
]{revtex4-2}
\makeatletter
\def\pre@bibdata{main3Notes}
\makeatother
\usepackage{amsmath}
\allowdisplaybreaks
\usepackage{amssymb}
\usepackage{graphicx}
\usepackage{longtable}
\usepackage{multirow}
\usepackage{epsfig}

\usepackage[usenames]{color}
\usepackage[
colorlinks=true,
linkcolor=blue,
citecolor=blue,
urlcolor=blue
]{hyperref}

\begin{document}

\title{Effects of Axion Interactions on Quark Stars in 4D Einstein--Gauss--Bonnet Gravity}

\author{Rui Zhou}
\affiliation{ %MOE Key Laboratory for Non-equilibrium Synthesis and Modulation of Condensed Matter, 
	School of Physics, Xi’an Jiaotong University, Xi’an 710049, China}
	
\author{Ming-zheng-xuan Wu}
	\affiliation{ %MOE Key Laboratory for Non-equilibrium Synthesis and Modulation of Condensed Matter, 
		School of Physics, Xi’an Jiaotong University, Xi’an 710049, China}
	
\author{Xin-ran Yang}
\affiliation{ %MOE Key Laboratory for Non-equilibrium Synthesis and Modulation of Condensed Matter, 
	School of Physics, Xi’an Jiaotong University, Xi’an 710049, China}

\author{Chong-long Xie}
\affiliation{ %MOE Key Laboratory for Non-equilibrium Synthesis and Modulation of Condensed Matter, 
	School of Physics, Xi’an Jiaotong University, Xi’an 710049, China}
	
\author{Min Zhou}
\affiliation{ %MOE Key Laboratory for Non-equilibrium Synthesis and Modulation of Condensed Matter, 
	School of Physics, Xi’an Jiaotong University, Xi’an 710049, China}

\author{Zhi-yang Liu}
\affiliation{ %MOE Key Laboratory for Non-equilibrium Synthesis and Modulation of Condensed Matter, 
	School of Physics, Xi’an Jiaotong University, Xi’an 710049, China}

\author{Shi-jun Mao}
\affiliation{ %MOE Key Laboratory for Non-equilibrium Synthesis and Modulation of Condensed Matter, 
	School of Physics, Xi’an Jiaotong University, Xi’an 710049, China}
				
\author{Guo-yun Shao}   
\email[Corresponding author ]{gyshao@mail.xjtu.edu.cn} 
%%\thanks{These authors contributed equally to this work}
\affiliation{ %MOE Key Laboratory for Non-equilibrium Synthesis and Modulation of Condensed Matter, 
School of Physics, Xi’an Jiaotong University, Xi’an 710049, China}

\begin{abstract}
	
We explore the properties of quark stars by combining the microscopic axion-extended Polyakov–Nambu–Jona-Lasinio model with the macroscopic framework of four-dimensional Einstein–Gauss–Bonnet (4D EGB) gravity. {Specifically, axion effects are incorporated through the background phase $\theta=a/f_a$ in the thermodynamics of quark matter.} Our results show that the inclusion of axion-induced interactions stiffens the quark-matter equation of state, thereby increasing the sound speed and the maximum mass of quark stars. The 4D EGB correction effectively weakens gravitational compression and further modifies the stellar structure, allowing for larger radii and higher maximum masses while reducing the compactness and the surface gravitational redshift. {The combined effects can support quark stars above $2\,M_\odot$, with mass--radius sequences overlapping the observational regions in the indicative phenomenological comparison.} These findings reveal that the interplay between axion-induced modifications of dense quark matter and 4D EGB gravity may provide a viable phenomenological framework for describing massive quark stars.

\end{abstract}

\maketitle

\section{Introduction}

Compact stars provide a unique laboratory for probing
strongly interacting matter under conditions of extreme density and gravity. One of the central open questions in this context
is the equation of state (EOS) of cold and dense matter,
which directly determines the global properties of compact
stars~\cite{PhysRevD.79.124032, LATTIMER2016127, ymsq-cfcw}. Understanding how new microscopic interactions and possible modifications
of gravity affect the EOS and stellar structure is therefore
of great importance.

Traditionally, compact stars have been modeled as objects
composed primarily of hadronic matter, while at higher
densities the appearance of deconfined quark matter may
lead to the formation of hybrid stars or even quark stars~\cite{Alford2007,PhysRevD.88.083013,Annala2020,Shao2011nu,Shao2013toa}.
However, it has also been suggested that compact stars
may contain nonbaryonic components beyond the standard
description of dense matter~\cite{SANDIN2009278, PhysRevD.103.043009, CIARCELLUTI201119}. In particular, if dark matter
accumulates efficiently inside compact stars through
gravitational capture or through its primordial abundance,
it may form an additional component that modifies the
internal composition, the EOS, and consequently the
observable mass-radius relation~\cite{KC2013, Guver_2014}.

Among the possible dark-matter candidates, the axion is
especially well motivated. It was originally proposed to solve the strong CP problem in quantum chromodynamics
(QCD)~\cite{PhysRevLett.38.1440,HSU2005369,Sannino2026wgx}. Owing to their weak interactions, cosmological stability, and nonthermal production through the vacuum misalignment mechanism, axions have long been regarded as promising candidates for cold dark matter~\cite{PRESKILL1983127, DINE1983137, ABBOTT1983133,PhysRevLett.40.223,PhysRevLett.40.279}.
Because of their extremely small mass and weak coupling to ordinary matter, axions are
also regarded as one of the most promising dark-matter candidates \cite{PhysRevD.80.035024,Duffy_2009},
and may even form Bose--Einstein condensates under suitable conditions
\cite{PhysRevLett.103.111301,PhysRevD.98.023009}. In recent years, the influence of axions
on the QCD phase structure \cite{PhysRevD.80.034019,PhysRevD.85.056009,PhysRevD.100.014013,
	PhysRevD.100.076021,PhysRevD.103.074003,PhysRevD.108.054010,
	PhysRevD.110.014042,Gong2024,PhysRevD.111.074001,6xx5-t88h}
and on compact-star properties
\cite{PhysRevD.106.L121301,PhysRevD.110.123031,PhysRevD.111.L051501,
	Lu2025,Kumar2026}
has attracted growing attention. In particular, axion interactions may modify the EOS of dense matter,
thereby affecting the macroscopic properties of compact stars.

On the other hand, despite the high-precision tests of Einstein's general relativity (GR) in the weak-field regime~\cite{Will2014}, the prevailing view is that GR is not the final theory of gravity.
Problems such as spacetime singularities \cite{PhysRevLett.14.57,10.1098/rspa.1970.0021},
non-renormalizability \cite{doi:10.1142/9789814539395-0001},
and several open issues in cosmology and astrophysics
\cite{Riess_1998,Perlmutter_1999,annurev:/content/journals/10.1146/annurev.astro.39.1.137}
have motivated extensive studies of modified gravity theories
\cite{APPLEBY20077,Ahmed2017-jj,10.1063/1.1665613}.
Among them, the four-dimensional Einstein--Gauss--Bonnet (4D EGB) gravity
\cite{d8c23c97-1e7d-30cf-af7c-c0828c0d0c95,ZWIEBACH1985315,WHEELER1986737,WILTSHIRE198636,PhysRevLett.124.081301}
has received particular attention as a higher-curvature extension of GR.
Although the Gauss--Bonnet term is topological in four dimensions in the standard formulation,
effective four-dimensional realizations have been proposed and have been  applied
to the study of compact objects. In particular, within the scalar-coupled formulation
of 4D EGB gravity, one can investigate the impact
of higher-curvature corrections on the structure of relativistic stars~\cite{LU2020135717,Banerjee_2021,TANGPHATI2021168498,PhysRevD.109.024026,PhysRevD.111.043034}. 
To date, studies of quark stars incorporating the interaction between axions and quark matter within the framework of 4D EGB gravity are still absent.

Motivated by these developments, we investigate quark stars within the framework of 4D EGB gravity in the presence of axion-induced effects. To this end, we extend the $(2+1)$-flavor Polyakov--Nambu--Jona-Lasinio (PNJL) model by incorporating vector interactions and a $\theta$-dependent 't Hooft term. {Our primary goal is to examine how axion effects modify the EOS and, together with the gravitational corrections from 4D EGB gravity, affect quark-star structure.} We compute the resulting EOS, mass--radius relations, compactness, and surface gravitational redshift, {and phenomenologically compare the predicted mass--radius sequences with observations.} Our results suggest that both the Gauss--Bonnet correction and the axion-induced quark interaction tend to increase the maximum mass of quark stars, albeit through distinct mechanisms. The former weakens gravitational compression, while the latter stiffens the EOS. {Their combined effects allow quark stars to reach masses comparable to those of observed massive pulsars in the phenomenological comparison, suggesting that the interplay between modified gravity and axion-modified dense matter can support massive quark-star configurations.}

This paper is organized as follows. In Sec.~\ref{sec:theory}, we briefly introduce the axion-extended PNJL model and the modified Tolman--Oppenheimer--Volkoff (TOV) equations in 4D EGB gravity. 
In Sec.~\ref{sec:results}, we present numerical results for the equation of state (EOS) and the mass--radius relations of quark stars under different conditions. 
Finally, we give our summary and conclusions in Sec.~\ref{sec:conclusions}.

\section{Axion-Extended PNJL Model and 4D EGB gravity}
\label{sec:theory}

\subsection{(2+1)-flavor PNJL model with axion interactions}

Building on Refs.~\cite{Lu2025,PhysRevD.106.L121301,PhysRevD.110.123031}, the Lagrangian density of the (2+1)-flavor PNJL model with axion interactions is given by
%\begin{equation}
  \begin{align}
  \mathcal{L} ={}& \bar{\psi}
  (i\gamma^\mu D_\mu+\gamma_0\hat{\mu}-\hat{m}_0)\psi \nonumber \\ 
  &-K \biggl\{ e^{i\theta}\det_f[\bar{\psi}(1+\gamma_5)\psi]+ e^{-i\theta}\det_f[\bar{\psi}(1-\gamma_5)\psi]\biggr\} \nonumber \\
  &-U(\Phi[A],\bar{\Phi}[A],T)+\mathcal{L}_S+\mathcal{L}_V
  \end{align} 
%\end{equation}
where $\psi$ denotes the quark fields of the three flavors, $u,\ d$, and $s$; $
\hat{m}_{0}=\mathrm{diag}(m_{u},\ m_{d},\ m_{s})$ in flavor space; 
$K$ is the six-point interaction constant.
 $\hat{\mu}=\mathrm{diag}(\mu_u,\mu_d,\mu_s)$ denotes the matrix of quark chemical potentials.
The covariant derivative is defined as $D_\mu=\partial_\mu-iA_\mu$.
The gluon background field $A_\mu=\delta_\mu^0A_0$ is supposed to be homogeneous and static, with  $A_0=g\mathcal{A}_0^\alpha \frac{\lambda^\alpha}{2}$, 
where $\lambda^\alpha/2$ are the generators of $SU(3)$ color.

Here, $\theta=a/f_a$ denotes the effective axion-induced CP-violating phase,
where $a$ is the axion field and $f_a$ is the axion decay constant. In the
underlying QCD theory, the physically relevant angular parameter is given by
the combination $\bar{\theta}+a/f_a$, where $\bar{\theta}$ denotes the strong
CP phase. Through the Peccei--Quinn mechanism, the axion field dynamically
relaxes the effective strong CP angle~\cite{PhysRevLett.38.1440,
	PhysRevLett.40.223,PhysRevLett.40.279}. In effective model studies, the
background $\bar{\theta}$ is therefore commonly absorbed into the axion
sector, and one simply writes $\theta\equiv a/f_a$.
In the PNJL model, this phase enters the instanton-induced 't Hooft interaction
through the factors $e^{\pm i\theta}$, thereby encoding the $U(1)_A$ anomaly
and introducing CP-odd effects into the effective quark interaction. As a
result, a nonzero $\theta$ induces mixing between scalar and pseudoscalar
channels and modifies the chiral and deconfinement properties of the
system~\cite{PhysRevD.85.056009,PhysRevD.103.074003,
	PhysRevD.108.054010}.
{ Since $\theta=a/f_a$ is treated as a fixed background phase in the present
	model, we assume that the axion field is static and spatially uniform throughout
	a quark star. Accordingly, each stellar sequence is constructed using the
	EOS evaluated at a fixed value of $\theta$.

The axion-induced interaction is $2\pi$-periodic in $\theta$, and the equilibrium thermodynamic observables are even under $\theta\to-\theta$. Consequently, the interval $0\leq\theta\leq\pi$ spans the independent angular range. The values $0$, $\pi/3$, $\pi/2$, $2\pi/3$, and $\pi$ sample this interval from the CP-conserving point to the endpoint associated with the Dashen mechanism~\cite{PhysRevD.108.054010,6xx5-t88h}. In particular, large values of $\theta$ probe the regime in which CP-odd effects and pseudoscalar condensates are most pronounced within the model~\cite{PhysRevD.80.034019}. These values are employed here as phenomenological probes of axion effects in dense quark matter.
\par}

The effective potential $U(\Phi[A],\bar{\Phi}[A],T)$ is expressed with the traced Polyakov loop
$\Phi=(\mathrm{Tr}_c L)/N_c$ and its conjugate
$\bar{\Phi}=(\mathrm{Tr}_c L^\dag)/N_c$. The Polyakov loop $L$  is a matrix in color space
\begin{equation}
	L(\vec{x})=\mathcal{P} \exp\bigg[i\int_0^\beta d\tau A_4 (\vec{x},\tau)   \bigg],
\end{equation}
where $\beta=1/T$ is the inverse temperature and $A_4=iA_0$. 
The Polyakov-loop effective potential taken in this study~\cite{roessner2007polyakov}   is
\begin{eqnarray} \label{U}
	\frac{U(\Phi,\bar{\Phi},T)}{T^4}&=&-\frac{a(T)}{2}\bar{\Phi}\Phi +b(T)\,\mathrm{ln}\big[1-6\bar{\Phi}\Phi\\ \nonumber
	&&+4(\bar{\Phi}^3+\Phi^3)-3(\bar{\Phi}\Phi)^2\big],
\end{eqnarray}
where
\begin{equation}
	\begin{aligned}
     a(T) =a_0+a_1\bigg(\frac{T_0}{T}\bigg)+a_2\bigg(\frac{T_0}{T}\bigg)^2, \,\,\,
  b(T) =b_3\bigg(\frac{T_0}{T}\bigg)^3,	
  \end{aligned} 
\end{equation}
with  $a_0=3.51$,  $a_1=-2.47$,  $a_2=15.2$, $b_3=-1.75$ and $T_0=210\,$ MeV.

The scalar and vector interactions parts are given by 

\begin{equation}
     \mathcal{L}_S = G_S \sum_{a=0}^8 \left[ (\bar{\psi} \lambda_a \psi)^2 
    + (\bar{\psi} i \gamma_5 \lambda_a \psi)^2 \right],	
\end{equation}
and 
\begin{equation}
  \mathcal{L}_V = -G_V \sum_{a=0}^8 \left[ (\bar{\psi} \gamma^\mu \lambda_a \psi)^2 + (\bar{\psi} \gamma^\mu \gamma_5 \lambda_a \psi)^2 \right],
\end{equation}
where $G_S$ and $G_V$ are the coupling constants of scalar and vector interactions, respectively. 

Within the mean-field approximation, the physical quark mass is given by
\begin{equation}
M_i=\sqrt{(M_i^s)^2+(M_i^p)^2},
\qquad i=u,d,s ,
\end{equation}
where $M_i^s$ and $M_i^p$ denote the scalar and pseudoscalar
contributions, respectively. They are
\begin{equation}
\begin{aligned}
M_i^s
=& m_{i0}-4G_S\sigma_i +2K\cos\theta\,(\sigma_j\sigma_k-\eta_j\eta_k)\\ 
& \qquad \qquad \quad-2 K \sin\theta\,(\sigma_j\eta_k+\sigma_k\eta_j)
\end{aligned}
\end{equation}
and
\begin{equation}
\begin{aligned}
M_i^p
=& 4G_S\eta_i 
+2K \cos\theta\,(\sigma_j\eta_k+\sigma_k\eta_j)\\
&\qquad \quad -2K \sin\theta\,(\sigma_j\sigma_k-\eta_j\eta_k).
\end{aligned}
\end{equation}
The scalar and pseudoscalar condensates are defined as
\begin{equation}
	\sigma_i=-\langle \bar{\psi}_i\psi_i\rangle,	\quad
	\eta_i=-\langle \bar{\psi}_i i\gamma_5\psi_i\rangle,
\end{equation}
and $(i,j,k)$ is any cyclic permutation of $(u,d,s)$.

The thermodynamic potential at finite temperature and chemical potential can be expressed as 
\begin{align}
\Omega ={}& \Omega_q+2G_S\sum_{i=u,d,s}(\sigma_i^2+\eta_i^2) \notag\\
&-4K\cos\theta\bigl(\sigma_u\sigma_d\sigma_s
-\sigma_u\eta_d\eta_s -\eta_u\sigma_d\eta_s-\eta_u\eta_d\sigma_s\bigr) \notag\\
&+4K\sin\theta\bigl(\eta_u\eta_d\eta_s
-\eta_u\sigma_d\sigma_s -\sigma_u\eta_d\sigma_s-\sigma_u\sigma_d\eta_s\bigr) \notag\\
&-2G_V(\rho_u^2+\rho_d^2+\rho_s^2)+U(\Phi,\bar{\Phi},T),
\end{align} 
where  $\rho_i$ is the quark number density of flavor $i$.
\begin{equation}
		\Omega_q= -2 N_c \sum_{i}\! \int^\Lambda \frac{d^3 p}{(2\pi)^3} 
		E_i \!-2 T\sum_{i}\! \int \frac{d^3 p}{(2\pi)^3} 
		[F^+_i + F_i^-]
\end{equation}
where the quasiparticle dispersion relation $E_i = \sqrt{\mathbf{p}^2 + M_i^2}$
and
\begin{equation}
  \begin{aligned}
F_i^{+} =& \ln\bigl[
  1 + 3\Phi\,\exp\!\bigl(-\tfrac{E_i-\tilde{\mu}_i}{T}\bigr)+ 3\bar{\Phi}\,\exp\!\bigl(-2\tfrac{E_i-\tilde{\mu}_i}{T}\bigr)
    \\ &
    + \exp\!\bigl(-3\tfrac{E_i-\tilde{\mu}_i}{T}\bigr)
\bigr], \\
F_i^{-} =& \ln\bigl[
  1 + 3\bar{\Phi}\,\exp\!\bigl(-\tfrac{E_i+\tilde{\mu}_i}{T}\bigr)+ 3\Phi\,\exp\!\bigl(-2\tfrac{E_i+\tilde{\mu}_i}{T}\bigr)
    \\ &
    + \exp\!\bigl(-3\tfrac{E_i+\tilde{\mu}_i}{T}\bigr)
\bigr].	
  \end{aligned} 
\end{equation}
Here  $\tilde{\mu}_i$ is the effective chemical potential defined as  $\tilde{\mu}_i=\mu_i-4G_V \rho_i.$

The self-consistently determined equations for the quark condensates and Polyakov
loop variables can be obtained by minimizing the thermodynamic potential,
\begin{equation}
   \frac{\partial \Omega}{\partial \sigma_i} =  \frac{\partial \Omega}{\partial \eta_i} 
   = \frac{\partial \Omega}{\partial \Phi} = \frac{\partial \Omega}{\partial \bar{\Phi}} =0.
\end{equation}
The pressure and energy density can be then obtained from
\begin{equation}
P = -(\Omega-\Omega_0),
\qquad
\epsilon = -P + TS+\sum_i \mu_i \rho_i,
\end{equation}
where $\Omega_0$ is the vacuum value of the thermodynamic potential, chosen such that $P=\epsilon=0$ in vacuum. A key feature of self-bound quark matter is that the pressure vanishes at the stellar surface while the energy density remains nonzero. To ensure this, a phenomenological vacuum pressure $B$ is usually introduced through $P\to P-B$ and $\epsilon\to\epsilon+B$. Following Ref.~\cite{27k9-r7nc}, we take $B^{1/4}=93$ MeV, which guarantees $\rho_B>\rho_0$ at the stellar surface.

For charge-neutral quark matter in $\beta$ equilibrium, the chemical potentials between different constituent species, including $u,d,s$ quarks and $e,\mu$ leptons, satisfy
\begin{equation}
\mu_d=\mu_u+\mu_e,\quad
\mu_s=\mu_d,\quad
\mu_\mu=\mu_e,
\end{equation}
subject to the charge-neutrality condition
\begin{equation}
\frac{2}{3}\rho_u-\frac{1}{3}\rho_d-\frac{1}{3}\rho_s-\rho_e-\rho_\mu=0.
\end{equation}
The leptons are treated as free Fermi gases. The particle number densities of electrons and muons are given by
\begin{equation}
	\rho_l = 2 \int \frac{d^3 p}{(2\pi)^3} 
	\left( \frac{1}{1+e^{(E_l-\mu_l)/T}} - \frac{1}{1+e^{(E_l+\mu_l)/T}} \right),
\end{equation}
where \(l\) denotes the lepton species (\(e\) or \(\mu\)).

\subsection{TOV equations in regularized 4D EGB gravity}
\label{sec:tov}

On the gravity side, we adopt the regularized  
4D EGB theory. It is worth noting that, although the Gauss--Bonnet term is purely
topological in exactly four spacetime dimensions, a consistent $D\to4$ formulation can be
obtained after regularization and can be recast as an effective scalar--tensor theory of
Horndeski type~\cite{LU2020135717,Kobayashi_2020, PhysRevD.102.024025}.
{ We adopt geometrized  units  $G=c=1$ for the gravitational equations.}
With this convention, the four-dimensional EGB action takes the form $S_{\rm EGB}= S_{\rm GR} + S^{(4)}_{\rm GB}$
~\cite{PhysRevD.109.024026}, where
\begin{equation}
S_{\rm GR}={\frac{1}{16\pi}}\int d^4x \sqrt{-g} R,
\end{equation}
\begin{equation}
\begin{aligned}
S^{(4)}_{\rm GB}
&={\frac{\alpha}{16\pi}}\int d^4x \sqrt{-g}
\bigl[\phi \mathcal{G}
+4G^{\mu\nu}\nabla_\mu\phi\nabla_\nu\phi \\
&\qquad-4(\nabla\phi)^2\Box\phi+2(\nabla\phi)^4\bigr].
\end{aligned}
\end{equation}
{The total bulk action is $S_{\rm tot}=S_{\rm EGB}+S_{\rm m}$,
where $S_{\rm m}$ is the minimally coupled matter action.}
The Einstein--Hilbert action \(S_{\mathrm{GR}}\) describes standard GR, while the regularized four-dimensional Gauss--Bonnet contribution \(S^{(4)}_{\mathrm{GB}}\) introduces higher-curvature corrections through an auxiliary dimensionless scalar field \(\phi\). 
$\mathcal{G}=R_{\mu\nu\rho\sigma}R^{\mu\nu\rho\sigma}$$
-4R_{\mu\nu}R^{\mu\nu}+R^2$ is the Gauss--Bonnet invariant and $\alpha$ is the Gauss--Bonnet coupling constant, which characterizes the strength of higher-curvature corrections. In the limit
$\alpha \to 0$, $S_{\rm EGB}$ reduces to $S_{\rm GR}$.

For a static, spherically symmetric compact star, we consider the metric
\begin{equation}
ds^2 = - e^{2\Phi(r)} dt^2 + e^{2\Lambda(r)} dr^2 + r^2 d\Omega^2 ,
\end{equation}
and  model stellar matter as a perfect fluid with energy-momentum tensor
\begin{equation}
T_{\mu\nu} = (\epsilon + P) u_\mu u_\nu + P g_{\mu\nu} .
\end{equation}
In standard GR, the \(g_{rr}\) component of the spherically symmetric metric, namely \(e^{2\Lambda(r)}\), is determined by the mass function \(M(r)\) through the relation
\begin{equation}
e^{-2\Lambda(r)} = 1 - \frac{2M(r)}{r}.
\end{equation}
{
In the regularized 4D EGB gravity, the scalar-field equation for the static,
spherically symmetric ansatz admits two branches~\cite{PhysRevD.109.024026},
\begin{equation}
\phi'_{\pm}(r)=[1\pm e^{\Lambda(r)}]/r.
\end{equation}
In an asymptotically flat spacetime, $e^{\Lambda(r)}=1+M/r+\mathcal{O}(r^{-2})$, so that $e^{\Lambda(r)}\to1$ as $r\to\infty$. For the plus branch, $\phi'_+(r)\sim2/r$, and $\phi_+(r)$ grows logarithmically, whereas the minus branch satisfies $\phi'_-(r)=-M/r^2+\mathcal{O}(r^{-3})$ and hence $\phi_-(r)=\phi_\infty+M/r+\mathcal{O}(r^{-2})$. Since asymptotic flatness requires $\phi(r)$ to approach a finite constant at infinity, we therefore adopt the minus branch. On this branch, introducing the mass function $M(r)$ and imposing asymptotic flatness, we obtain
\par}
\begin{equation}
	e^{-2\Lambda(r)} = 1 + \frac{r^2}{2\alpha}
	\left(
	1 - \sqrt{1 + \frac{8\alpha M(r)}{r^3}}
	\right),
\end{equation}
{
	where the minus sign in front of the square root selects the branch connected to general relativity, since $e^{-2\Lambda(r)}=1-2M(r)/r+\mathcal{O}(\alpha)$ as $\alpha\to0$.
	\par}
{On this branch}, the modified TOV equations can be written as

\begin{equation}
\label{eq:TOV_EGB}
\frac{dP}{dr}
=\frac{(\epsilon+P)\left[r^3\left(\Gamma+8\pi\alpha P-1\right)-2\alpha M\right]}
{r^2\Gamma\left[r^2(\Gamma-1)-2\alpha\right]},
\end{equation}
and
\begin{equation}
\label{eq:mass_EGB}
\frac{dM}{dr}=4\pi r^2 \epsilon .
\end{equation}

 The quantity \(\Gamma\) appearing in  the modified TOV equations is defined as
 \begin{equation}
 	\Gamma=\sqrt{1+\frac{8\alpha M}{r^3}},
 \end{equation}
 and originates from the algebraic solution of the \(g_{rr}\) component of the field equations in regularized 4D EGB gravity. Physically, the parameter $\alpha$ in \(\Gamma\) serves as a convenient measure of the deviation from general relativity, which captures the effect of the Gauss–Bonnet correction on the gravitational potential.
 In the limit \(\alpha\to 0\), one has \(\Gamma\to 1\), and Eqs.~\eqref{eq:TOV_EGB} and \eqref{eq:mass_EGB}
 reduce to the standard TOV equations in GR.

In the calculation, the boundary conditions are taken as
\begin{equation}
M(r=0)=0,\qquad P(r=0)=P_c,
\end{equation}
where  $P_c$ is the central pressure in quark stars, and the stellar radius $R$ is defined by the condition
\begin{equation}
P(r=R)=0.
\end{equation}
{
	For $r>R$, the asymptotically flat vacuum solution has
	$M(r)=M(R)\equiv M$ and $e^{2\Phi}=e^{-2\Lambda}=f_{\rm ext}(r)$, where
	\begin{equation}
		f_{\rm ext}(r)=1+\frac{r^2}{2\alpha}
		\left(1-\sqrt{1+\frac{8\alpha M}{r^3}}\right).
	\end{equation}
	At $P(R)=0$, the interior metric and scalar field are matched continuously to the asymptotically flat exterior solution on the same GR-connected branch, so that the induced metric, extrinsic curvature, scalar field, and its normal derivative are continuous. Although the surface energy density $\epsilon(R)$ remains finite, the mass function is continuous, and no surface stress-energy layer or localized scalar source is introduced.
	\par}

\subsection{Model parameters and numerical setup}

In this work, we employ the $(2+1)$-flavor PNJL parameter set as 
 $\Lambda = 630.0$ MeV, $m_u = m_d = 5.5$ MeV, $m_s = 135.7$ MeV, $G_S \Lambda^2 = 1.781$, and $K \Lambda^5 = 9.29$~\cite{rehberg1996hadronization}.
We fix the vector coupling at $G_V=0.5\,G_S$, consistent with recent studies of compact-star properties \cite{Shao2013toa,PhysRevD.110.123031}.
The Gauss--Bonnet coupling constant $\alpha$ is treated as a parameter, but constrained by astrophysical observations to be $\alpha \le 10^2 \, \mathrm{km}^2$~\cite{PhysRevD.102.084005}. The present calculations focus on cold quark stars in the zero-temperature limit.

\section{Numerical results and discussion}
\label{sec:results}

\subsection{Equation of state of quark matter with axion interaction}

\begin{figure}[htb]
	\begin{center}
		\includegraphics[scale=0.38]{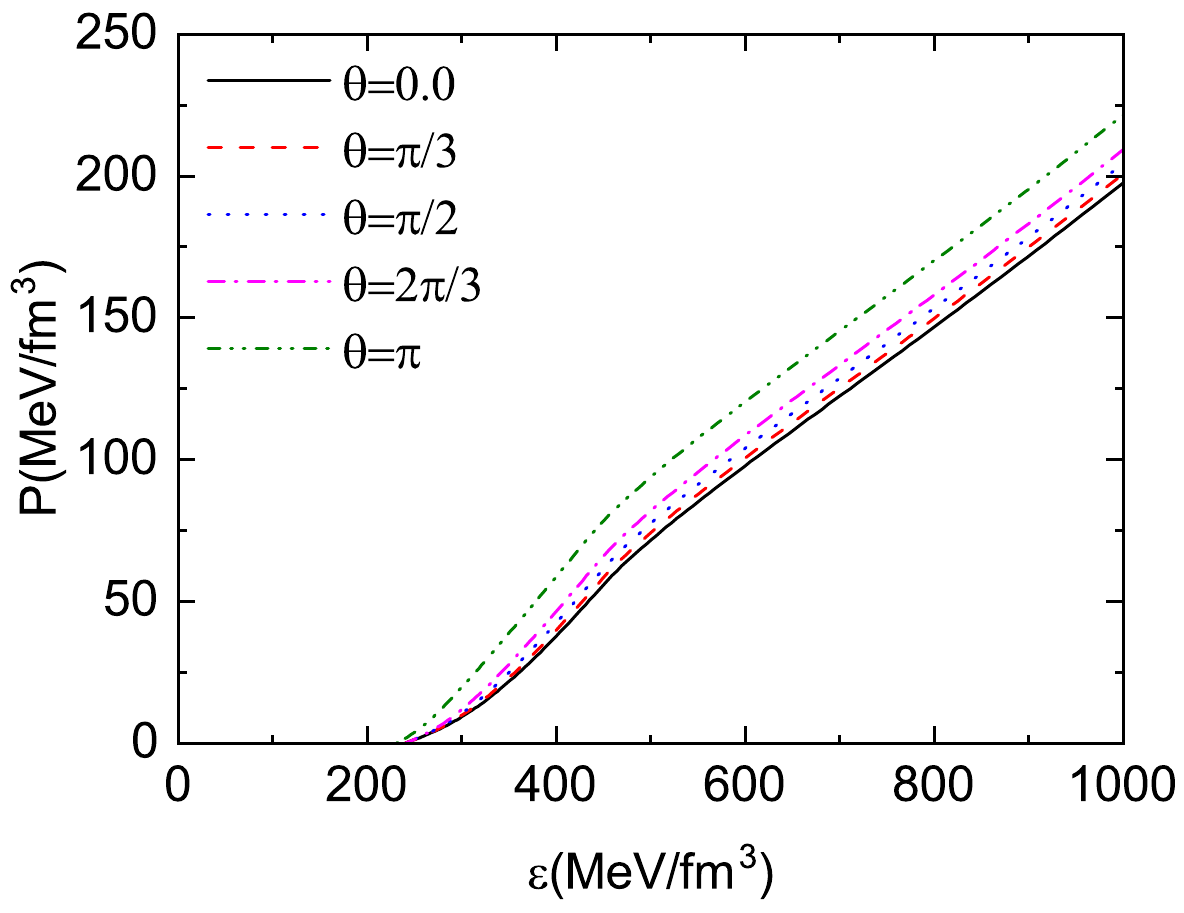}
		\caption{\label{fig:EOS_axion}{
			Equation of state of quark matter for
			$\theta=0$, $\pi/3$, $\pi/2$, $2\pi/3$, and $\pi$.}}
	\end{center}
\end{figure}

First, we present  in Fig.~\ref{fig:EOS_axion} the EOSs of quark matter with the axion interaction for $\theta=0,\pi/3,\pi/2,2\pi/3,$ and $\pi$.
{
	When constructing the EOS of quark matter with axion interactions, $\Lambda$,
	$m_u$, $m_d$, $m_s$, $G_S$, $K$, $G_V$, $B$, and $T=0$ are fixed at the values
	specified above, and only $\theta$ is treated as a parameter. For each value of $\theta$ 
	and given chemical potentials, we solve the gap equations in Eq.~(14)
	self-consistently using Eqs.~(8)--(11).
 The gap equations
	are solved together with the $\beta$-equilibrium and charge-neutrality
	conditions, and the resulting solutions are used to derive the pressure $(P)$ and energy density $(\epsilon)$.
}
In all cases, Fig.~\ref{fig:EOS_axion} shows that the pressure increases monotonically with energy density, and the energy density remains finite at zero pressure, reflecting the nonvanishing surface density characteristic of self-bound quark matter. Moreover, at fixed energy density, the pressure rises systematically with increasing $\theta$, which indicates that the axion interaction stiffens the quark-matter EOS within the present PNJL framework with vector interactions.

\begin{figure}[htb]
	\begin{center}
		\includegraphics[scale=0.38]{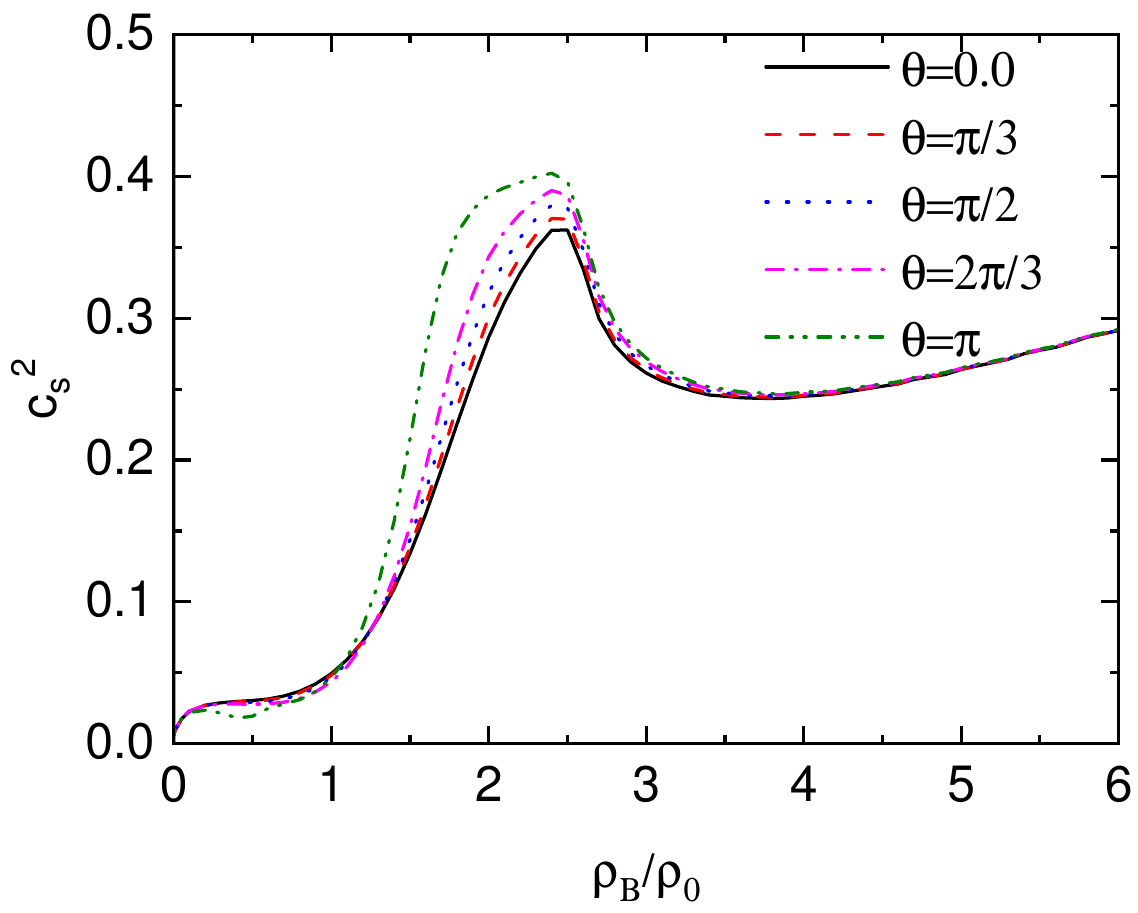}
		\caption{\label{fig:EOS_cs}  Squared speed of sound as a function 
			of  $\rho_B/\rho_0$ with axion interactions. }
	\end{center}
\end{figure}

\begin{figure}[htb]
\begin{center}
	\includegraphics[scale=0.38]{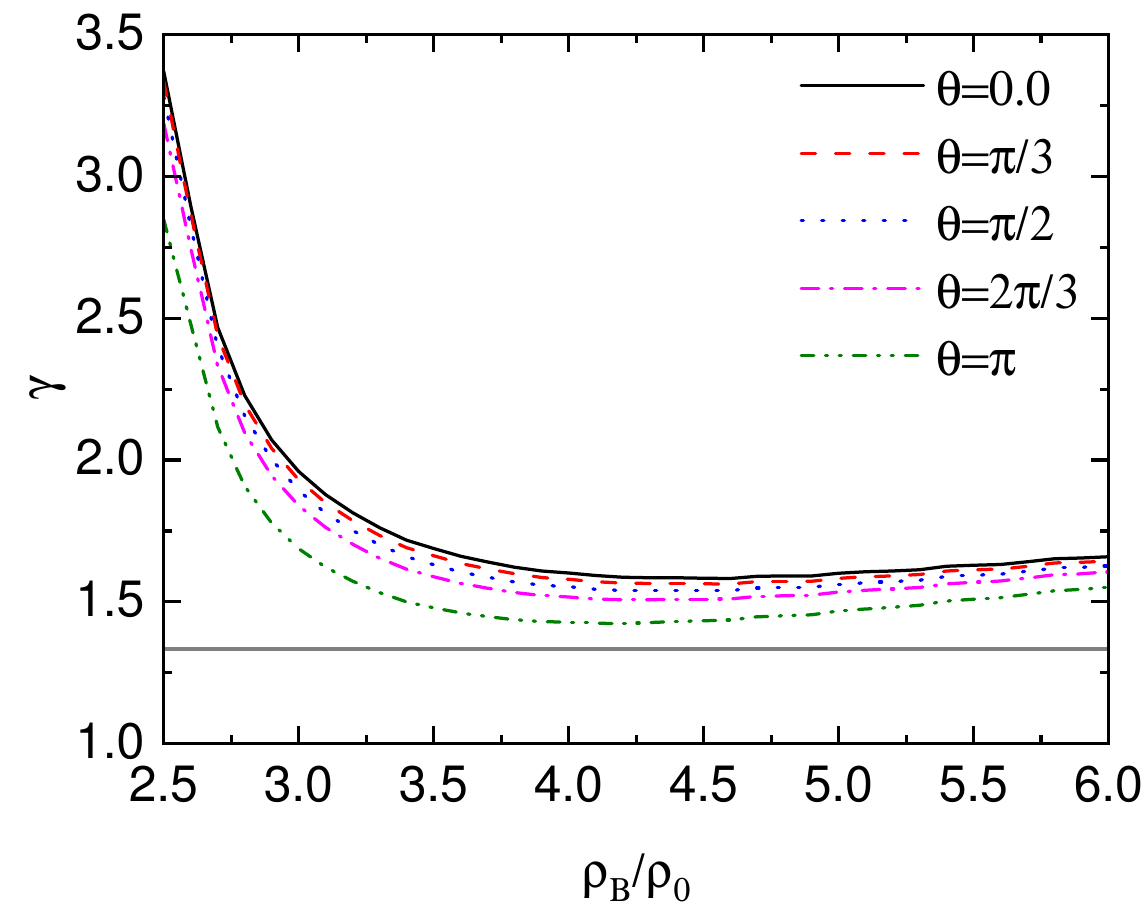}
	\caption{\label{fig:EOS_gamma}  Adiabatic index as a function 
			of  $\rho_B/\rho_0$ with axion interactions.
			{The horizontal line denotes the Newtonian reference value $\gamma=4/3$.}}
\end{center}
\end{figure}

To further quantify the influence of axion interactions on the EOS,
we show the squared sound speed, $c_s^2=(dP/d\epsilon)_s$, and the adiabatic index,
$\gamma=(1+\epsilon/P)\, (dP/d\epsilon)_s$, in
Figs.~\ref{fig:EOS_cs} and \ref{fig:EOS_gamma}, respectively.
As shown in Fig.~\ref{fig:EOS_cs}, the sound speed exhibits a pronounced peak
at intermediate densities for all considered values of $\theta$.
While the peak position changes mildly, its height increases noticeably
with increasing $\theta$, indicating that the axion interaction enhances the
stiffness of quark matter in this density region.
For $\rho_B>2.5\rho_0$, $c_s^2$ decreases with density for all values of $\theta$.
This behavior can be attributed to the appearance of strange quarks at higher densities, which tends to soften the EOS. At very high densities, these curves gradually increase again and converge for different values of $\theta$. A similar behavior is also observed in quark matter in the absence of beta-equilibrium conditions~\cite{PhysRevD105094024}.


The adiabatic index $\gamma$ characterizes the local stiffness
of quark matter EOS. As seen in Fig.~\ref{fig:EOS_gamma}, increasing $\theta$ leads to a moderate reduction of $\gamma$,
which appears opposite to the behavior of $c_s^2$.
This can be understood from the relation
$\gamma=(1+\epsilon/P)c_s^2$.
Although the axion interaction increases $c_s^2$, it also raises the pressure
at fixed energy density, thereby reducing the prefactor $(1+\epsilon/P)$.
The combined effect  is a suppression of $\gamma$.
{
The adiabatic index $\gamma$ is related to the response of stellar matter to adiabatic radial perturbations. In Newtonian gravity, $\gamma=4/3$ is commonly used as a reference critical value for radial stability. In GR, the critical value of $\gamma$ is larger than $4/3$, and its precise value depends on the global stellar structure~\cite{Moustakidis2017}. In this work, we do not rigorously compute the critical value of $\gamma$ in the presence of 4D EGB gravity corrections. Figure~\ref{fig:EOS_gamma} provides only the variation of $\gamma$ with $\theta$, which is insufficient to fully demonstrate relativistic radial stability. A more detailed investigation of this issue is left for future work.
}

\subsection{Stellar structure of quark stars in 4D EGB gravity}

\begin{figure}[htb]
	\begin{center}
		\includegraphics[width=0.4\linewidth]{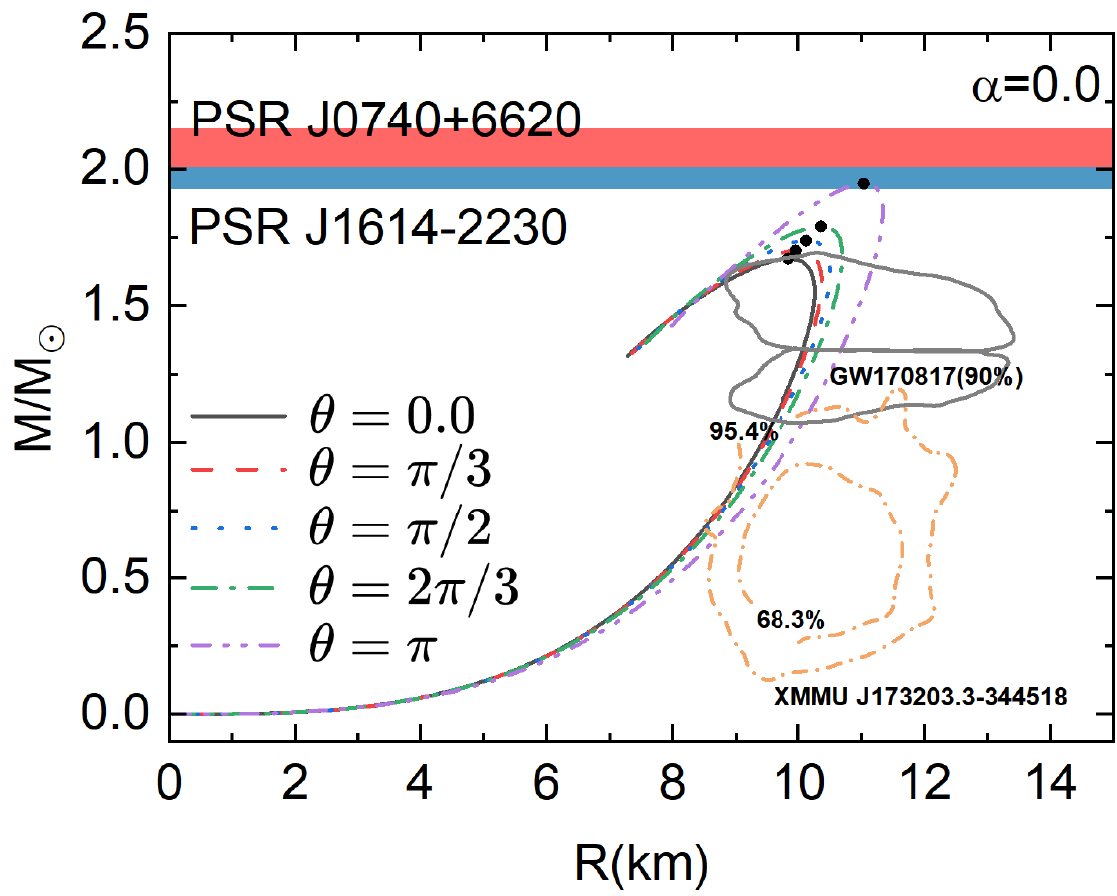}
		\caption{\label{fig:RM_axion}  Mass-radius relations of quark stars with axion interactions
			in Einstein gravity.
			{The red and blue horizontal bands denote
				$M/M_\odot\in[2.01,2.15]$ for PSR J0740+6620
				($2.08\pm0.07\,M_\odot$, 68.3\%)~\cite{Fonseca2021} and
				$M/M_\odot\in[1.93,2.01]$ for PSR J1614-2230
				($1.97\pm0.04\,M_\odot$, $1\sigma$)~\cite{Demorest2010},
				respectively. The gray contours show the 90\% credible regions
				for the two components of GW170817~\cite{PhysRevLett.119.161101,PhysRevLett.121.161101}; the orange inner and outer
				contours show the 68.3\% and 95.4\% regions for
				XMMU J173203.3-344518~\cite{Doroshenko2022}, respectively.}}
	\end{center}
\end{figure}

Figure~\ref{fig:RM_axion} presents the mass-radius relations of quark stars for different values of $\theta$ within the standard GR framework (with $\alpha=0$). Maximum-mass configurations are marked with black dots. As $\theta$ increases, the stellar sequences shift toward higher masses
and slightly larger radii, reflecting the axion-induced stiffening of the
quark-matter EOS. In particular, the maximum mass increases from
$1.67\,M_\odot$ at $\theta=0$ to $1.94\,M_\odot$ at $\theta=\pi$,
while the corresponding radius increases from $9.8$ km to $11.0$ km.
This indicates that the axion interaction enhances not only the maximum
supported mass but also the characteristic size of the quark star.

\begin{figure}[htb]
	\begin{center}
		\includegraphics[width=0.4\linewidth]{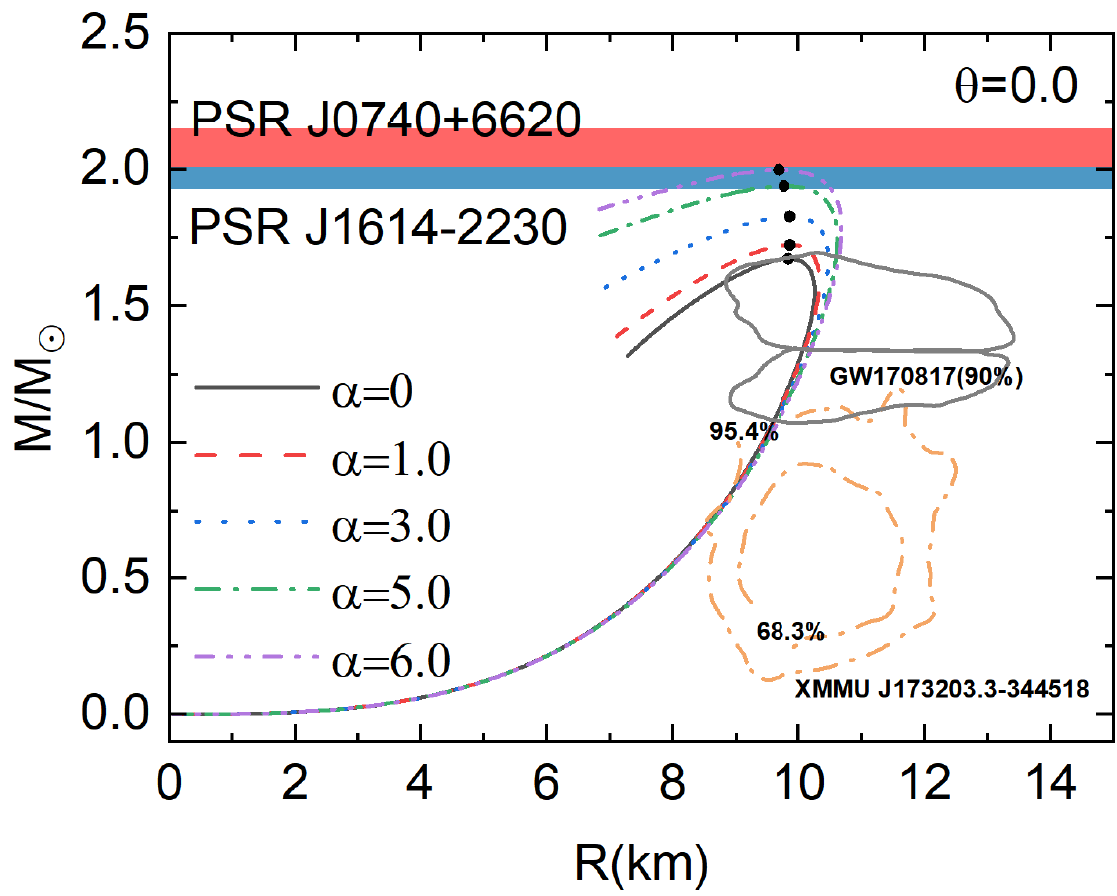}
		\caption{\label{fig:RM_alpha}  
			Mass-radius relations of quark stars
			with different values of $\alpha$ at fixed $\theta=0$.
            {The observational bands and contours are the
            same as in Fig.~\ref{fig:RM_axion}.}}
	\end{center}
\end{figure}

\begin{figure}[htbp]
	\begin{center}
		\includegraphics[width=0.4\linewidth]{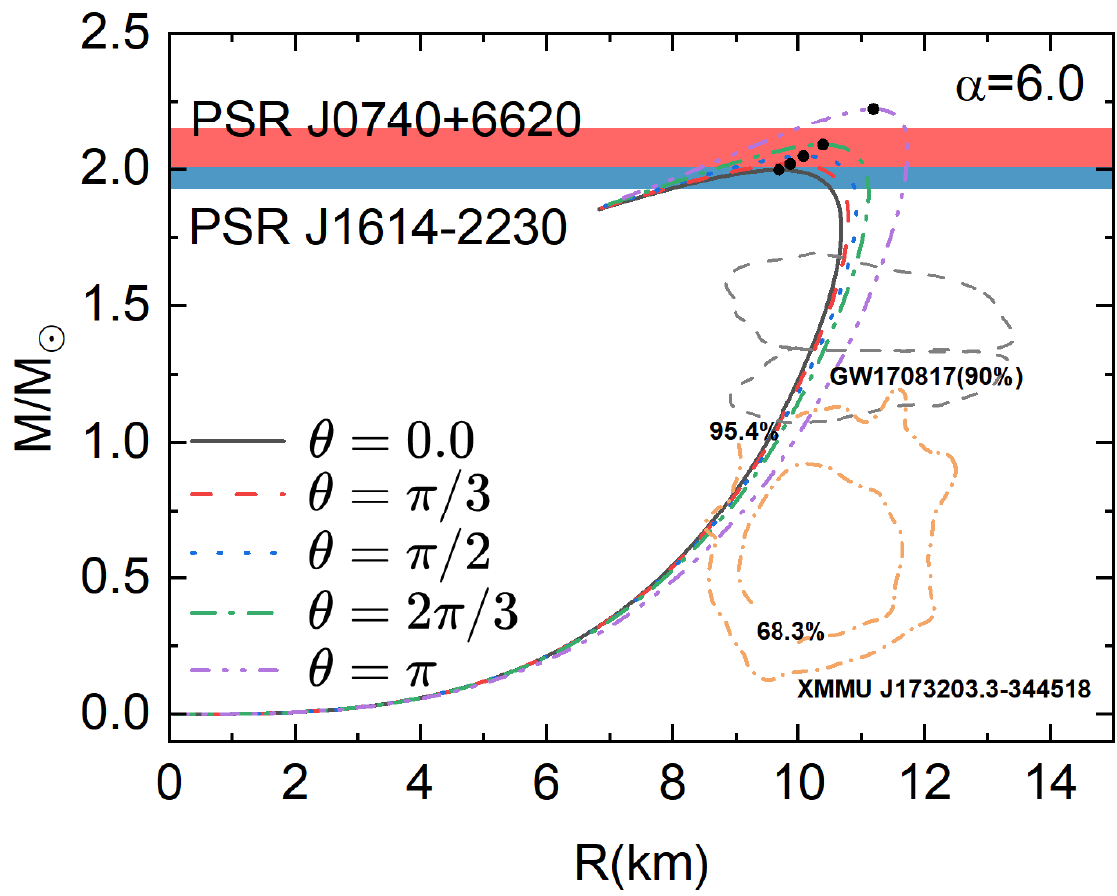}
		\caption{\label{fig:RM_final}  Mass-radius relations of quark stars with axion interactions
			in 4D EGB gravity with $\alpha=6.0 \, \mathrm{km}^2$.
			{The observational bands and contours are the
				same as in Fig.~\ref{fig:RM_axion}.}}
	\end{center}
\end{figure}

{
For comparison with observations, Figs.~\ref{fig:RM_axion}, \ref{fig:RM_alpha}, and
\ref{fig:RM_final} also include the masses of
PSR J0740+6620~\cite{Fonseca2021} and PSR J1614-2230~\cite{Demorest2010},
together with the mass--radius regions inferred from
GW170817~\cite{PhysRevLett.119.161101,PhysRevLett.121.161101} and
XMMU J173203.3-344518~\cite{Doroshenko2022}.
The GW170817 regions were inferred from tidal-deformability data within GR,
assuming a common EOS and low spins for the two neutron stars. The XMMU
J173203.3-344518 regions were obtained within GR from thermal X-ray spectra
using a uniform carbon-atmosphere model and a Gaia-based distance. Their
overlay on the $\alpha\neq0$ sequences is intended as an
indicative phenomenological comparison.
At $\alpha=0$, the maximum mass reaches the lower edge of the
PSR J1614-2230 mass band, $1.93\,M_\odot$, only for $\theta=\pi$.
\par}

Next, we investigate the influence of Gauss--Bonnet coupling constant $\alpha$ on the mass--radius relation. Observational constraints from binary black holes suggest an upper bound on 
this coupling, $\alpha \lesssim 10^2\,\mathrm{km}^2$ \cite{PhysRevD.102.084005}.
In this work, we consider the representative values
$\alpha = 0,\,1.0,\,3.0,\,5.0,$ and $6.0\,\mathrm{km}^2$,
which are also consistent with recent studies of compact stars in 4D EGB gravity
\cite{TANGPHATI2021168498,Pretel2022}.

The corresponding mass-radius relations,
obtained at fixed $\theta=0$, are displayed in Fig.~\ref{fig:RM_alpha}. As $\alpha$ increases, the stellar sequences shift toward larger radii and slightly
higher masses. For a fixed stellar mass, the radius becomes larger,
while the maximum mass also increases monotonically with $\alpha$, as indicated by
the black dots in Fig.~\ref{fig:RM_alpha}. These results
suggest that the 4D EGB correction effectively weakens gravitational
compression, thereby supporting more massive and less compact quark-star configurations.

\begin{figure}[htb]
	\centering
	\includegraphics[width=0.4\linewidth]{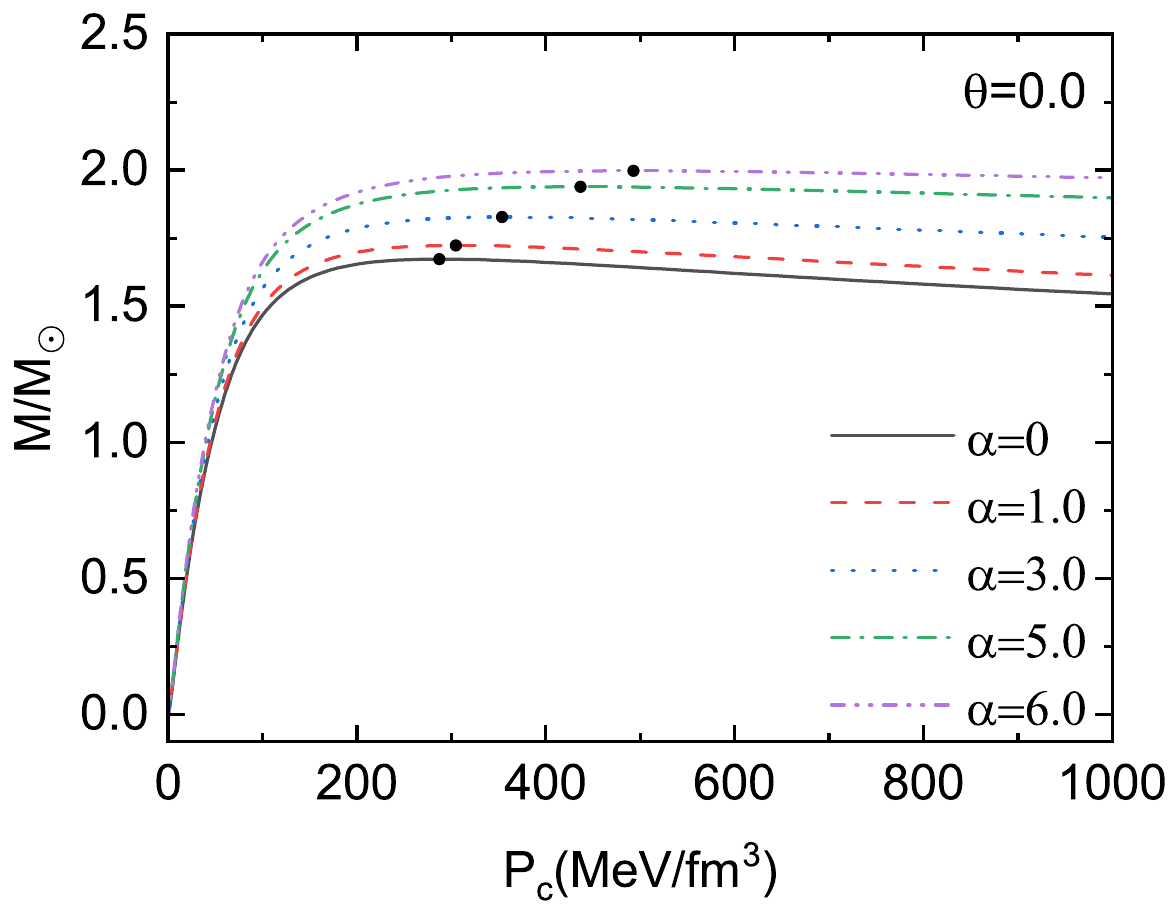} 
	\vspace{0.0cm} % 调整两张图片之间的垂直间距
	\includegraphics[width=0.4\linewidth]{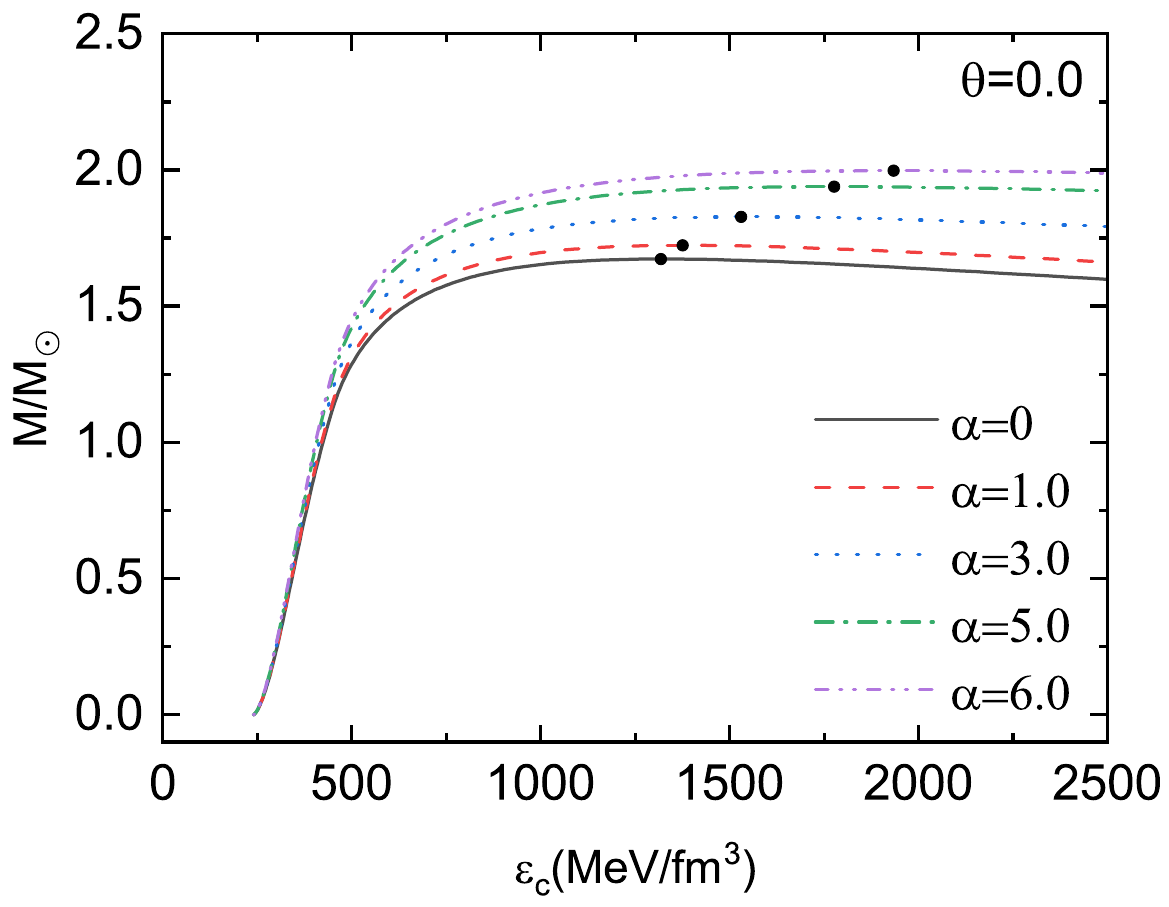}
	\caption{\label{fig:P_e_c_alpha} 
		Central pressure (left panel) and central energy density (right panel) as 
		functions of stellar mass with different values of $\alpha$.
		{Black dots mark the maximum-mass configurations.}}
\end{figure}

To better understand how the Gauss--Bonnet coupling affects the internal stellar
structure, we show the central pressure $P_c$ and central energy density $\epsilon_c$
as functions of stellar mass in Fig.~\ref{fig:P_e_c_alpha}.
{
In GR, the mass extremum along a one-parameter sequence of cold stars
marks the onset of fundamental radial instability~\cite{Bardeen1966}. For the
regularized 4D EGB branch considered here, direct radial-mode calculations for
several neutron-star EOSs indicate that this coincidence is retained
~\cite{PhysRevD.111.064071}. Here we adopt $dM/d\epsilon_c=0$
(equivalently $dM/dP_c=0$) as an approximate turning-point indicator for the
quark-star sequences. Accordingly, the black dots in
Fig.~\ref{fig:P_e_c_alpha} indicate the expected onset of instability, and the
branches on either side are hereafter referred to as the expected stable and
unstable branches, respectively.
\par}

For a given central pressure or central energy density, the supported stellar
mass increases with increasing $\alpha$. Equivalently, for a fixed stellar mass
on the stable branch, a larger value of $\alpha$ generally requires a lower
central pressure and energy density to maintain hydrostatic equilibrium,
reflecting the effective weakening of gravitational compression.  Figure~\ref{fig:P_e_c_alpha} also
shows that increasing
$\alpha$ shifts the maximum-mass configuration toward higher
$M_{\max}$, $P_c$, and $\epsilon_c$.

\begin{figure}[htb]
	\centering
	\includegraphics[width=0.4\linewidth]{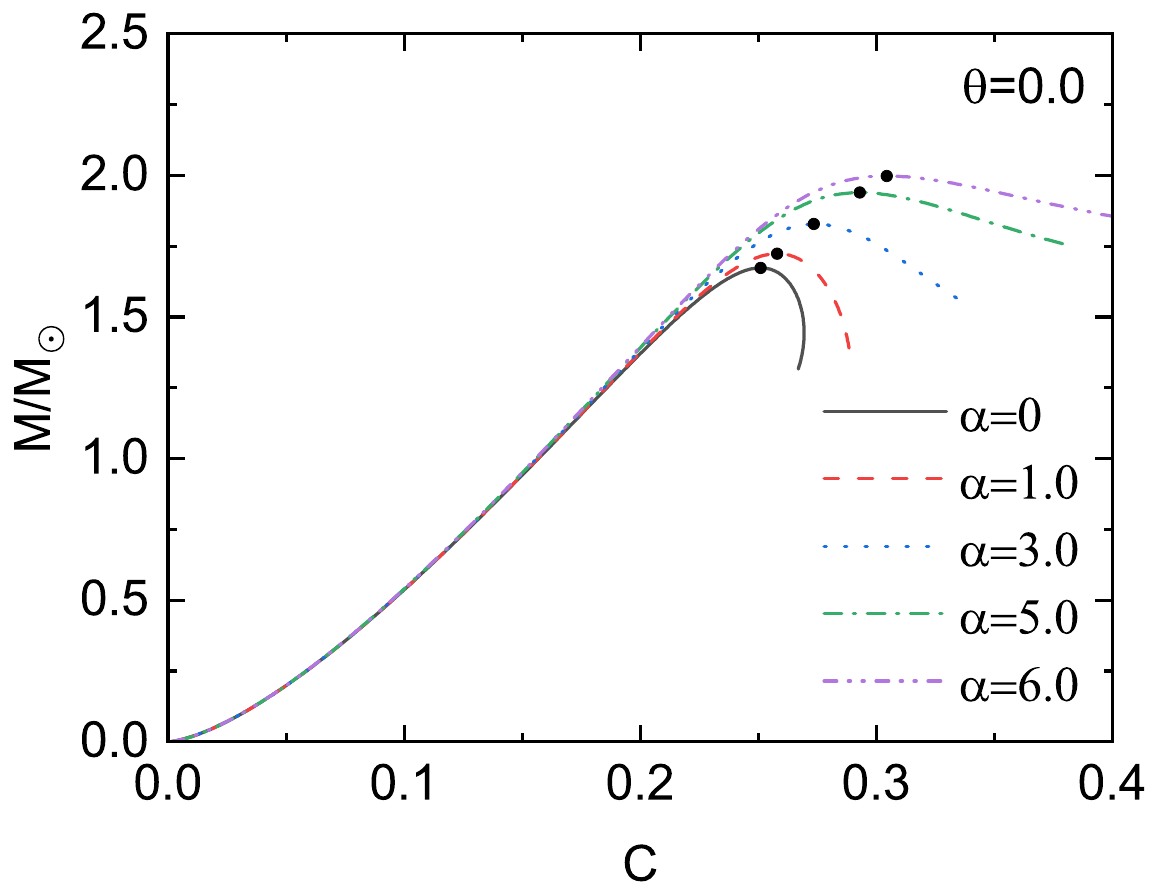} 
	\vspace{0.0cm} % 调整两张图之间的上下间距
	\includegraphics[width=0.4\linewidth]{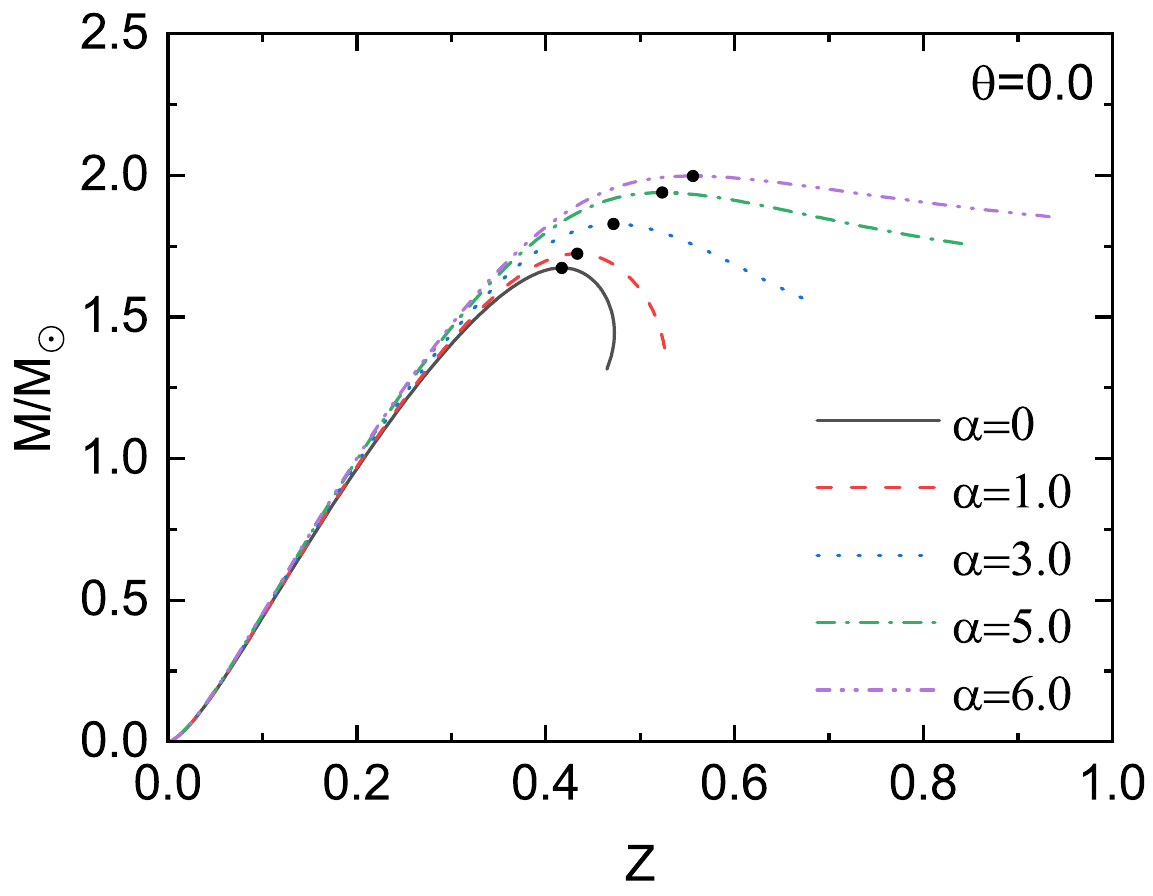}   
	\caption{\label{fig:C_Z_alpha} 
		Compactness (left panel) and surface redshift (right panel) as 
		functions of stellar mass with different values of $\alpha$.}
\end{figure}

We further examine the compactness, $C=M/R$, and the surface gravitational
redshift in 4D EGB gravity. {With the exterior solution
specified in Sec.~\ref{sec:tov}, the redshift measured by a static observer at
infinity is $Z=f_{\rm ext}(R)^{-1/2}-1$, which yields}
\begin{equation}
Z =
\left[
1+\frac{R^2}{2\alpha}
\left(1-\sqrt{1+\frac{8\alpha M}{R^3}}\right)
\right]^{-1/2}-1 .
\end{equation}
For $\alpha=0$, this expression is understood in the general-relativistic limit.

Figure~\ref{fig:C_Z_alpha} presents the compactness (left panel) and surface redshift (right panel) as functions of stellar mass with different values of $\alpha$.  
In the stable branch, both $C$ and $Z$
decrease monotonically with increasing $\alpha$ at fixed stellar mass. This trend
is fully consistent with the mass-radius relations in Fig.~\ref{fig:RM_alpha}.
Although the EGB correction increases the maximum supported mass, it also leads
to larger stellar radii, so that the overall compactness of quark stars is reduced.
Since the surface redshift is intimately tied to the gravitational potential at the stellar surface, a smaller compactness necessarily implies a smaller value of 
$Z$. Therefore, the 4D EGB correction does not merely shift the maximum mass upward; it also modifies the global stellar structure by favoring configurations that are less compact and, consequently, less gravitationally redshifted than their counterparts in standard GR.

Figure~\ref{fig:RM_final} shows the mass-radius relations of quark stars
with axion interactions in 4D EGB gravity for $\alpha=6.0\,\mathrm{km}^2$.
Compared with the corresponding results in Einstein gravity shown in
Fig.~\ref{fig:RM_axion}, all stellar sequences are shifted toward higher
masses and slightly larger radii. This demonstrates that the axion-induced stiffening of the microscopic EOS and the EGB correction in the stellar structure equations jointly modify the global properties of quark stars.

{
For all considered values of $\theta$, the quark-star sequences overlap with the observational constraints from GW170817 and XMMU J173203.3-344518 in the phenomenological comparison. Their maximum masses exceed the lower bound of the PSR J1614-2230 mass band, $1.93\,M_\odot$. For $\theta \geq \pi/3$, they also exceed the lower bound of the PSR J0740+6620 mass band, $2.01\,M_\odot$. These results suggest that the combined effects of axion interactions and 4D EGB gravity could improve the compatibility of quark-star configurations with current observations.
\par}

\begin{figure}[htb]
	\centering
	\includegraphics[width=0.4\linewidth]{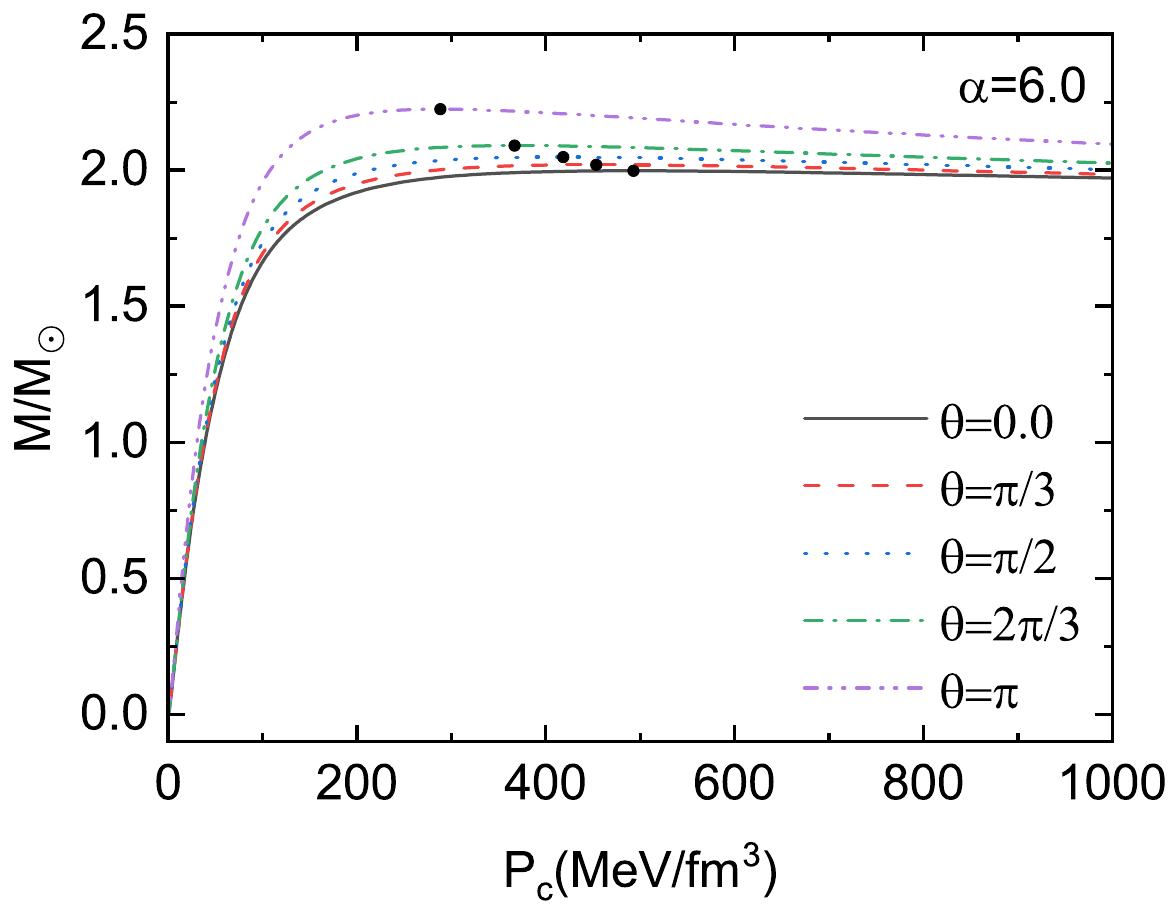} 
	\includegraphics[width=0.4\linewidth]{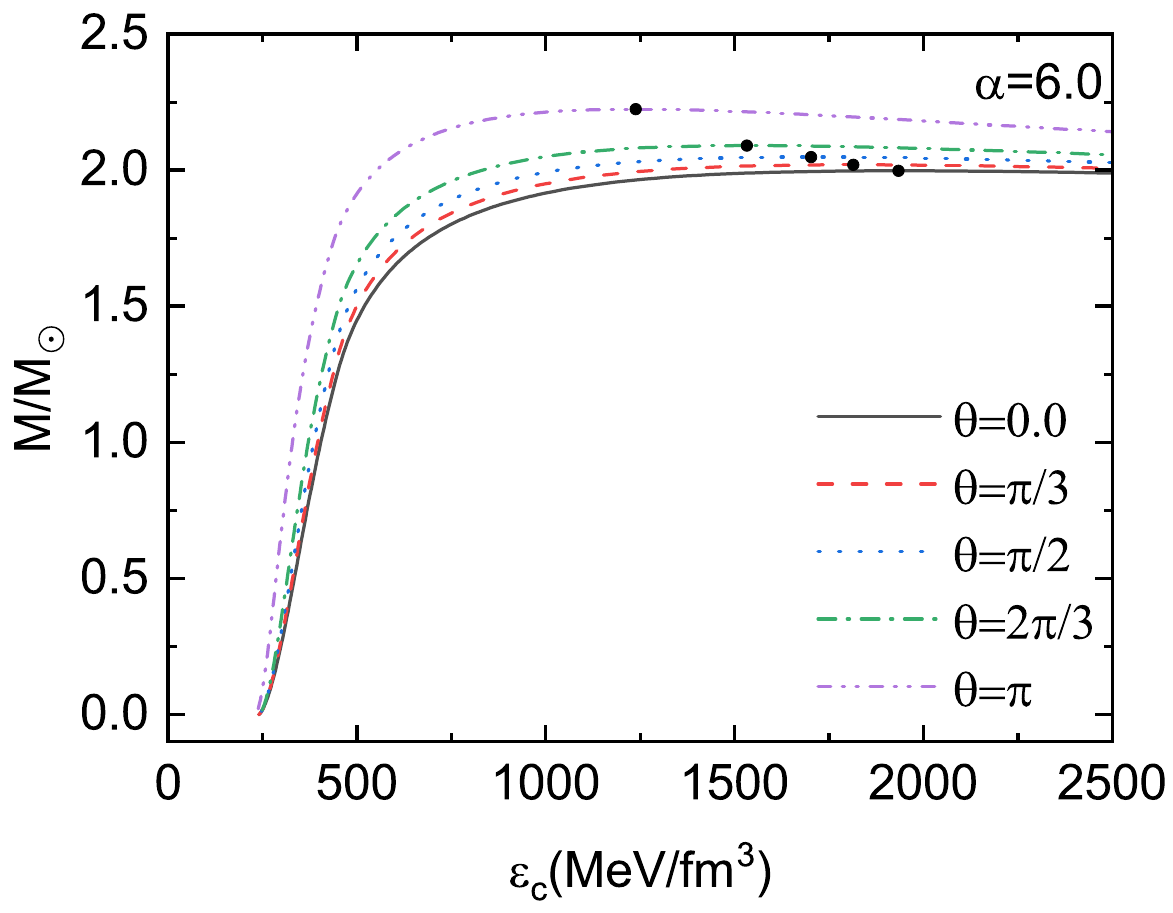}    
	\caption{\label{fig:P_e_c_final} 
    Central pressure (left panel) and central energy density (right panel) as 
    functions of stellar mass with axion interactions.}
\end{figure}

We further analyze the central pressure $P_c$ and central energy density
$\epsilon_c$ as functions of stellar mass in Fig.~\ref{fig:P_e_c_final}.
For a given central pressure or central energy density, a larger value of
$\theta$ corresponds to a larger stellar mass, which is consistent with
the axion-induced stiffening of the EOS.
In particular,
the expected turning point of the equilibrium sequence shifts toward
lower central energy densities as $\theta$ increases. This indicates
that, in the presence of stronger axion effects, a more massive
quark star can be supported at a relatively lower central energy
density (central pressure). In other words, the enhanced effective repulsion in the EOS provides greater support against gravitational collapse, allowing the system to remain in hydrostatic equilibrium up to higher masses and shifting the onset of instability to larger masses.

\begin{figure}[t]
	\centering
	\includegraphics[width=0.4\linewidth]{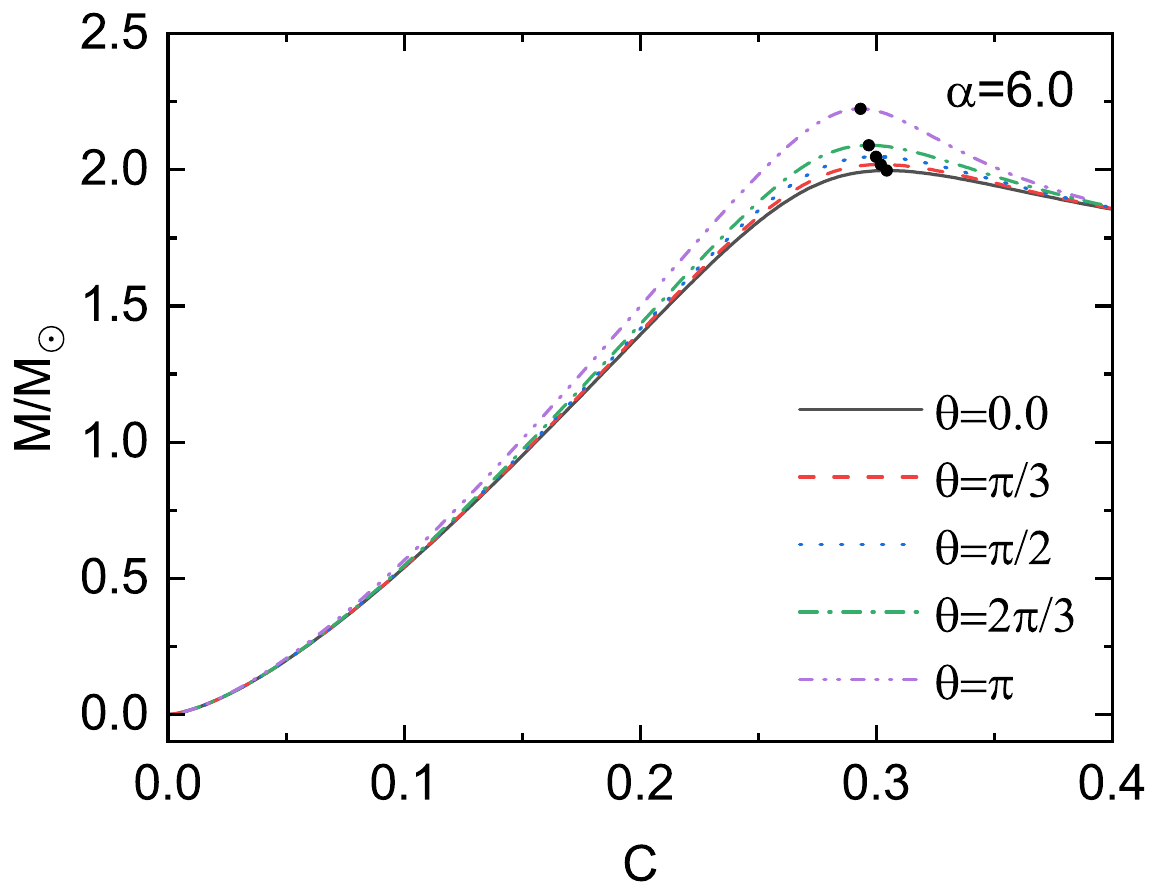} 
	\includegraphics[width=0.4\linewidth]{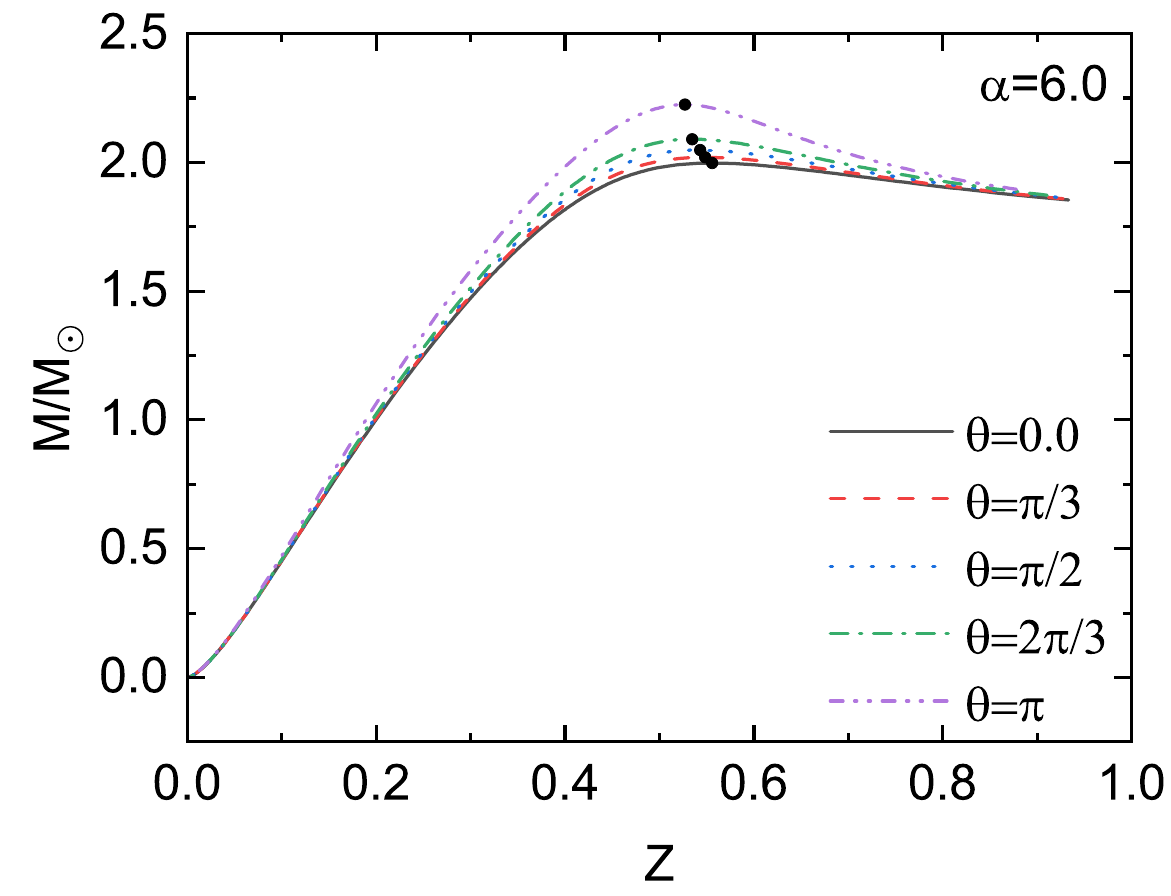}  
	\caption{\label{fig:C_Z_final}
    Compactness (left panel) and surface redshift (right panel) as
    functions of stellar mass for different values of $\theta$.}
\end{figure}

The compactness $C$ and the surface redshift $Z$ as functions
of stellar mass for different values of $\theta$ are shown in
Fig.~\ref{fig:C_Z_final}. In the stable branch, both $C$ and $Z$ decrease
monotonically with increasing $\theta$ at fixed stellar mass. This trend is
a direct consequence of the axion-induced stiffening of the EOS, which leads
to larger stellar radii and therefore less compact configurations. In the unstable branch, however, the curves for different $\theta$ values
become much closer to each other.  
This tendency suggests that the structural properties of the star are increasingly governed by
the ultra-dense quark matter background at extremely high density, while the relative impact of 
axion interactions becomes less pronounced.

\section{Summary and conclusions}
\label{sec:conclusions}
In this study, we have investigated the effects of regularized 4D EGB gravity and axion interactions on the properties of quark stars within the PNJL model. By varying the Gauss--Bonnet coupling constant $\alpha$ and the axion parameter $\theta$, we studied the interplay between macroscopic gravitational corrections and microscopic modifications of dense quark matter.

We found that increasing the Gauss--Bonnet coupling constant $\alpha$ enhances the maximum mass of quark stars. This behavior can be understood as an effective weakening of gravitational compression induced by the EGB correction,  {which allows quark stars to support higher-mass equilibrium configurations.} At the same time, for a fixed stellar mass, a larger value of $\alpha$ leads to a larger radius, and therefore to a smaller compactness and a lower surface gravitational redshift. This indicates that 4D EGB gravity favors more massive but less compact quark-star configurations than those in standard general relativity.

The inclusion of axion interactions significantly stiffens the equation of state of quark matter.
The increase in $\theta$ leads to a larger pressure at fixed energy density, a higher sound speed, and consequently a greater maximum mass of quark stars. We also found that the maximum-mass configuration shifts toward lower central energy densities as $\theta$ increases, indicating that the axion-induced stiffening enables the system to maintain hydrostatic equilibrium at relatively lower central densities.

By combining the effects of 4D EGB gravity and axion interactions, we showed that the mass-radius relations of quark stars can be brought into better agreement with {observations in the phenomenological comparison}. In particular, for representative parameter choices such as $\alpha = 6.0\,\mathrm{km}^2$ and $\theta \ge \pi/3$, the resulting quark-star sequences can {reach the lower edge of the PSR J0740+6620 mass band and overlap the mass--radius regions inferred from GW170817 and XMMU J173203.3-344518}. These results suggest that the combined effects of modified gravity and axion-induced stiffening of the equation of state could provide a viable mechanism for sustaining massive quark stars.

\begin{acknowledgments}
This work is supported by the National Natural Science Foundation of China under
Grant No.~12475145, 12275204  and Natural Science Basic Research Plan in Shaanxi Province
of China (Program No.~2024JC-YBMS-018).

\end{acknowledgments}

\bibliographystyle{apsrev4-2}
\bibliography{apssamp}

\end{document}